\documentclass[aps,prb,10pt]{revtex4-2}
\usepackage{subfig}
\usepackage{graphicx}% Include figure files
\usepackage{epsfig}
\usepackage{dcolumn}% Align table columns on decimal point
\usepackage{bm}% bold math
\usepackage{times}
\usepackage{nicematrix}
\usepackage{amsmath}

\begin{document}
\title{Berry-phase effect in single molecule magnets: analytical and numerical results}
\author{Fco.\ Javier Anaya-Garc\'ia}
\address{Coordinaci\'on para la Innovaci\'on y la Aplicaci\'on de la Ciencia y la Tecnolog\'ia, Universidad Aut\'onoma de San Luis Potos\'i, San Luis Potos\'i, 78000 MEXICO}
\author{Daniel Salgado-Blanco}
\address{C\'atedra CONACYT--Centro Nacional de Superc\'omputo, Instituto Potosino de Investigaci\'on Cient\'ifica y Tecnol\'ogica, Camino a la Presa San Jos\'e 2055, 78216, San Luis Potos\'i, MEXICO}
\author{Gabriel Gonz\'alez}
\address{C\'atedra CONACYT--Universidad Aut\'onoma de San Luis Potos\'i, San Luis Potos\'i, 78000 MEXICO}
\address{Coordinaci\'on para la Innovaci\'on y la Aplicaci\'on de la Ciencia y la Tecnolog\'ia, Universidad Aut\'onoma de San Luis Potos\'i, San Luis Potos\'i, 78000 MEXICO}
\address{NanoScience Technology Center, Department of Physics, University of Central Florida, Orlando, FL 32826, USA}
\email{gabriel.gonzalez@uaslp.mx}
\begin{abstract}
 In this paper  we theoretically and numerically investigate transport signatures of quantum interference on the current through a single molecule magnet transistor tunnel coupled to oppositely polarized leads in the presence of a local transverse and longitudinal magnetic field. Our calculations are based in a density matrix approach where we treat the ground state energy splitting induced by tunneling of the spin between different paths with the aid of perturbation theory. Using this approach we show that it is possible to use an effective Hamiltonian which describes the Berry phase interference as a function of the transverse magnetic field which completely blocks the current flow when we place the single molecule magnet between oppositely polarized leads. Finally, we use this effective Hamiltonian in an open source Python software (QmeQ) that allows us to calculate the current through the single molecule magnet with oppositely polarized leads tunnel coupled to the single molecule magnet. The analytical results are well reproduced by our numerical simulations.
\end{abstract}
%%%%%%%%%%%%%%%%%%%%%%%%%%%%%%%%%%%%%%%%%%%%%%%%%%%%%%%%%%%%%%%%%%%%%



\maketitle

%%%%%%%%%%%%%%%%%%%%%%%%%%%%%%%%%%%%%%%%%%%%%%%%%%%%%%%%%%%%%%%%%%%%%%%%%%%%%%
\section{Introduction}
Quantum mechanical interference effects are interesting phenomena that have a wide range of applications in science and technology. Since the first experiments in single molecule magnets revealed evidence of quantum spin tunneling at low temperatures and Berry phase interference effects in the presence of a transverse magnetic field (which leads to the vanishing of the energy splitting \cite{wernsdorfer2002, leuenberger2006, friedman2010single,thoss2018perspective}), single molecule magnets (SMM), such as Mn (see Refs. \cite{quddusi2011,yao2020berry}) and Fe (see Refs. \cite{bruno2004,bianco2014,yao2020berry}), have been a topic of growing interest in nanoscience.
Single-molecule magnets (SMMs) are mesoscopic systems that present a "giant" spin at low temperatures
and that exhibit uniquely quantum dynamics such as magnetic quantum tunneling and quantum interference effects. The Berry phase effect has been studied theoretically by the path-integral method and perturbative approaches and has also been detected experimentally in several single molecule magnets by measuring their magnetization dynamics. The finger print of the Berry phase effect in spin molecule systems is determined by the suppression of tunnel splitting when different spin tunneling paths interfere destructively for certain values of the transverse magnetic field\cite{Wrzesniewski2017,Wernsdorfer1999,Garg1993}. \\
More recently, electron transport across a three-terminal SMM has been explored in various studies, and the Coulomb blockade has been demonstrated. The Kondo effect oscillations as a function of the transverse magnetic field has also been studied \cite{gonzalez2008kondo}.
The Berry-phase interference effect in electron transport in an SMM transistor with opposite spin polarized leads was also investigated theoretically \cite{gonzalez2007berry}. Electron transport features of a SMM transistor were largely studied theoretically with a constant gate voltage and, more recently, the Floquet formalism was used to tackle the problem of electron transport over a periodically driven ferromagnetic quantum barrier with a local magnetic field and ac-driven potential \cite{gonzalez2019berry}. \\

In this work we study the Berry-phase effect in electron transport through an SMM under a local longitudinal and transverse magnetic field connected to oppositely polarized leads.
We show analytically and numerically that the current oscillates as a function of the transverse magnetic field due to the Berry phase interference associated with two quantum tunneling paths. The article is organized as follows. In
the first section we will give a detailed theoretical description of the Berry phase effect from a perturbative point of view.  Using this approach we show that it is possible to use an effective Hamiltonian which describes the Berry phase interference as a function of the transverse magnetic field which completely blocks the current flow when we place the single molecule magnet between oppositely polarized leads. In the second section, we use this effective Hamiltonian and use it in a open source Python software (QmeQ)\cite{qmeq2017} that allows us to calculate the current through the single molecule magnet with oppositely polarized leads tunnel coupled to the single molecule magnet and show that the theoretical results are in very good agreement with the numerical calculations. The conclusions are summarized in the last section.

\section{Density Matrix: Analytical Formulation}
\begin{figure}[htbp]
	    \centering
	    \includegraphics[scale=0.75]{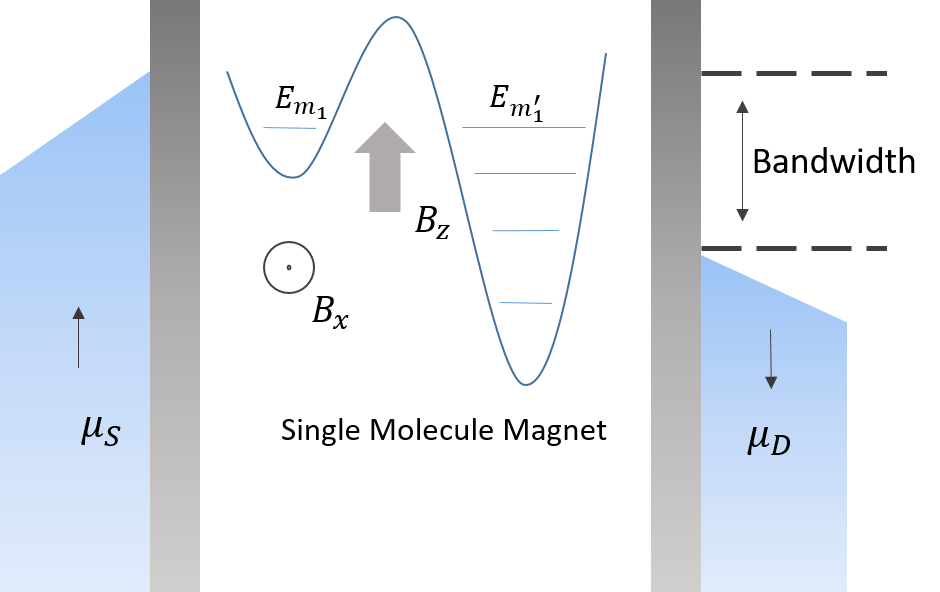}
		\caption{SMM in the Coulomb blocked regime connected to source and drain contacts, under the influence parallel and transverse magnetic field. One can see the resonant states $E_{m_1}$ and $E_{m_1^{'}}$ of the SMM}.
	    \label{fig:smm4}
\end{figure}

Adding a term for a longitudinal magnetic field to the simplest effective Hamiltonian which describes the Berry-phase of a spin quantum tunneling in SMMs, given by \cite{gonzalez2019berry, Romero_2014}, we obtain
\begin{equation}
{\cal H}^{(q)}=-K_zS_{q,z}^2-g\mu_B\left(B_zS_{q,z}+B_xS_{q,x}\right)+K(S_{q,x}^2-S_{q,y}^2)+J_{\parallel}S_{q,z}(n_{\uparrow}-n_{\downarrow}) +J_{\perp}(S_{q}^{+}c^{\dagger}_{\downarrow}c_{\uparrow}+S_{q}^{-}c^{\dagger}_{\uparrow}c_{\downarrow}),
\label{eq01}
\end{equation}
where ${\cal H}^{(q)}$ denotes the energy of the SMM for the un-charged ($q=0$) and charged ($q=1$) case, $K_z$ is the uniaxial anisotropy constant, $B_z$ and $B_x$ is the local longitudinal and transverse magnetic field, $K$ is the in-plane transverse anisotropy constant, $S_{q,x}$, $S_{q,y}$ and $S_{q,z}$ are quantum charged ($q=1$) or un-charged ($q=0$) molecule spin operators, $n_{\sigma}=c^{\dagger}_{\sigma}c_{\sigma}$ is the electron number operator where $\sigma$ denotes the electron's spin, $S_{q}^{\pm}=S_{q,x}\pm iS_{q,y}$ are molecule spin-flip operators and $J_{\parallel}$ and $J_{\perp}$ are the longitudinal and transverse exchange interactions. We will consider that the exchange coupling between the electron and the SMM is highly anisotropic so that the spin-flip between the electron and the SMM is suppressed, which means that $J_{\perp}=0$. \cite{gonzalez2019berry,timm2008tunneling,misiorny2009switching}
We work in the strong Coulomb blockade regime such that there is at most one excess electron in the SMM, we also assume that the SMM is coupled to oppositely spin polarized leads (See Fig. (\ref{fig:smm4})). \\
We will now assume that $B_x$ and $K$ in equation (\ref{eq01}) are small compared to $K_z$, $B_z$ and $J_{\parallel}$ so that we can rewrite the Hamiltonian given in equation (\ref{eq01}) in the following form
\begin{equation}\label{eq02}
{\cal H}^{(q)}={\cal H}_0+\delta{\cal H}
\end{equation}
where the "unperturbed" Hamiltonian ${\cal H}_0$ is given by
\begin{equation}\label{eq03}
{\cal H}_0=-K_zS_{q,z}^2-g\mu_BB_zS_{q,z} +J_{\parallel}S_{q,z}(n_{\uparrow}-n_{\downarrow})
\end{equation}
while the perturbation $\delta{\cal H}$ is given by
\begin{equation}\label{eq04}
\delta{\cal H}=-g\mu_BB_xS_{q,x}+K(S_{q,x}^2-S_{q,y}^2)
\end{equation}
It is easy to see that the energies for ${\cal H}_0$ when $q=1$ are given as a function of $S_{1,z}=m_1$ by
\begin{equation}\label{eq05}
E_{m_1,\sigma}^{(0)}=-K_zm_1^2-g\mu_BB_z m_1 +J_{\parallel} m_1\left(\delta_{\sigma,\uparrow}-\delta_{\sigma,\downarrow}\right)
\end{equation}
Since we have oppositely polarized leads we can assume without loss of generality that the incoming electrons hoping into the SMM have a positive spin, i.e. $\sigma=\uparrow$, which means that we can tune the longitudinal magnetic field $B_z$ in such a way to have a resonant tunneling between two different spin projections $m_1$ and $m_1^{\prime}$, where we will assume that $m_1>0$ and $m_1^{\prime}<0$, respectively. The degeneracy between $m_1$ and $m_1^{\prime}$ occurs if
\begin{equation}\label{eq06}
  E_{m_1,\uparrow}^{(0)}-E_{m_1^{\prime},\downarrow}^{(0)}=0
\end{equation}
The condition given by equation (\ref{eq06}) implies that we will have degeneracy for the charged SMM when a longitudinal magnetic field is given by
\begin{equation}\label{eq07}
  g\mu_BB_z=-K_z(m_1+m_1^{\prime})+J_{\parallel} \left(\frac{m_1-|m_1^{\prime}|}{m_1+|m_1^{\prime}|}\right)
\end{equation}
%Due to the indluence of $B_z$, $m_1 = s-n$ and $m_1^{\prime} = -s$, with $n = 1,2,3,...$ small. Since we are dealing with SMM's, $2s>>n$, and we are able to define an ${K_z}^{(eff)}=K_z\left(1-\dfrac{J_\parallel}{2sK_z}\right)$, in a way that
%\begin{equation}
 %   \dfrac{B_z}{K_z^{(eff)}}=n
%\end{equation}
%and equation (\ref{eq05}) can be written as
%\begin{equation}
%E_{m_1,\sigma}^{(0)}=-K_z m_1^2 -m_1B_z + \dfrac{J_\parallel n}{2s}
%\label{levelcross}
%\end{equation}
%For a fixed spin, we get a group of straight lines by graphing $B_z$ against $E_{m_1,\sigma}^{(0)}$. We found degeneration in the energy levels, consequence of the presence of the parallel magnetic field. It is represented by the intersections between the lines associated to different energy levels shown in Figure \ref{fig:smm5}.
%\begin{figure}[htbp]
%	    \centering
%	    \includegraphics[scale=0.65]{EvsB}
%		\caption{Crossing between different spin levels in equation \ref{levelcross}, for a SMM with spin $s = 9/2$}
%	    \label{fig:smm5}
%\end{figure}
Therefore, when the longitudinal magnetic field fulfills equation (\ref{eq07}) the unperturbed Hamiltonian ${\cal H}_0$ has a diagonal matrix form given by
\begin{equation}\label{eq08}
{\cal H}_0=
\begin{pNiceMatrix}[first-row,last-row,first-col,last-col,nullify-dots]
& |s_1,m_1\rangle &|-s_1,m_1^{\prime}\rangle & \\
\langle s_1,m_1|& E_{m_1,\uparrow}^{(0)} & 0  \\
 \langle -s_1,m_1^{\prime}|   & 0 & E_{m_1^{\prime},\downarrow}^{(0)}  \\
  & &
\end{pNiceMatrix}
\end{equation}
where $| s_1,m_1\rangle$ and $| -s_1,m_1^{\prime}\rangle$ denotes the charged molecular eigenstates.
The degeneracies in the Hamiltonian (\ref{eq08}) can be removed by the perturbation $\delta{\cal H}$ which do not commute with $S_{1,z}$. The energy splitting induced by the quantum tunneling of the spin can be obtained by applying perturbation theory to the Hamiltonian ${\cal H}_0$, which is given by \cite{Krizanac_2017,Garg2017,yoo2005}
\begin{equation}
\label{eq09}
\Delta E=\frac{4s}{2^{2s}(2s-1)!K_z^{2s-1}}\prod_{n=1}^{2s}\left(B_x-(2s+1-2n)B_a\right),
\end{equation}
where $B_x$ and $K$ represents the two perturbations, $s$ represents the ground spin state of the SMM and $B_a=\sqrt{2KK_z}$. 

\begin{figure}[htbp]
	    \centering
	    \includegraphics[scale=1.3]{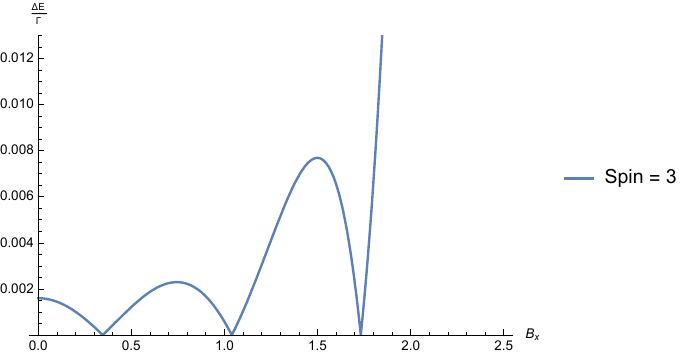}
		\caption{The energy splitting obtained as a result of destructive quantum destructive interference between two different spin tunneling paths. This effect is also known as Berry's phase interference. The parameters of the system are  $s = 3$, $K = 0.06$ meV, $K_z = 1$ meV and $\Gamma = 0.38$.}
	    \label{fig:Berry}
\end{figure}

From Eq. (\ref{eq09}) we see that the tunnel splitting is zero when $B_x^{(n)}=(2s+1-2n)B_a$ for $n=1,2,\ldots,2s$. The vanishing of the tunnel splitting for certain values of the transverse magnetic field is due to the destructive interference between different spin tunneling paths.  This is called the the Berry's phase interference. In Fig. \ref{fig:Berry} is possible to see this  effect in a SMM with spin $s=3$. Therefore, we end up with an effective two level Hamiltonian for the system which is given by
\begin{equation}\label{eq082}
{\cal H}^{(q=1)}=
\begin{pNiceMatrix}[first-row,last-row,first-col,last-col,nullify-dots]
&|s_1,m_1\rangle &|-s_1,m_1^{\prime}\rangle & \\
\langle s_1,m_1| & E_{m_1,\uparrow}^{(0)} & \Delta E  \\
 \langle -s_1,m_1^{\prime}|  & \Delta E & E_{m_1^{\prime},\downarrow}^{(0)}  \\
  & &
\end{pNiceMatrix}
\end{equation}

The density matrix $\rho (t)$ for the system contains entries
not only for charged spin states $|s_1,m_1\rangle$, but also
for the uncharged spin state $|s_0,m_0\rangle$. The time evolution of the density matrix is given by the generalized master equation in the Lindblad form within the Born-Markov and infinite-bias approximations.

\begin{equation}
    \dfrac{d\hat{\rho}(t)}{dt} = -i \left[ {\cal H}^{(q=1)}, \hat{\rho}(t)\right] + \displaystyle\sum_i \left( \hat{D}_i \hat{\rho}(t) \hat{D}_i^\dagger - \dfrac{1}{2}\left\lbrace \hat{D}_i^\dagger \hat{D}_i,\hat{\rho}(t) \right\rbrace \right) \equiv \hat{\mathcal{L}} \hat{\rho}(t)
    \label{eq:Lindblad}
\end{equation}

\noindent where the jump operators $\hat{D}_i$ describe irreversible tunneling
of electrons into and out of the system with rates $\Gamma_i$, and are given by:

\begin{equation}
\begin{matrix}
    D_1 & = \sqrt{\Gamma_1}|s_1,m_1 \rangle \langle s_0,m_0 |\\
    D_2 & = \sqrt{\Gamma_2}|s_0,m_0 \rangle \langle -s_1,m_1^{\prime} |
\end{matrix}.
\label{eq:salto}
\end{equation}

\noindent Taking $\Gamma_1=\Gamma_2=\Gamma$, the Lindblad equation for this system reads:
\begin{equation}
    \dfrac{d\rho}{dt}= -i[{\cal H}^{(q=1)},\rho] + D_1\rho D_1^\dagger + D_2\rho D_2^\dagger - \dfrac{1}{2} \left( D_1^\dagger D_1 \rho + D_2^\dagger D_2 \rho  \right)
     - \dfrac{1}{2} \left( \rho D_1^\dagger D_1 + \rho D_2^\dagger D_2  \right)
     =\mathcal{L} \rho .
     \label{eq:lindblad_smm}
\end{equation}

\noindent The stationary properties of the system are determined by the eigenvalues and eigenvectors of $\mathcal{L}$. The stationary desity matrix $\rho (\infty)$ is given by the eigenvector of $\mathcal{L}$ with zero eigenvalue, therefore the stationary average current is given by $\langle I \rangle = \Gamma \rho_{33}(\infty)$.

\begin{figure}[htbp]
	   \centering
	    \includegraphics[scale=1.2]{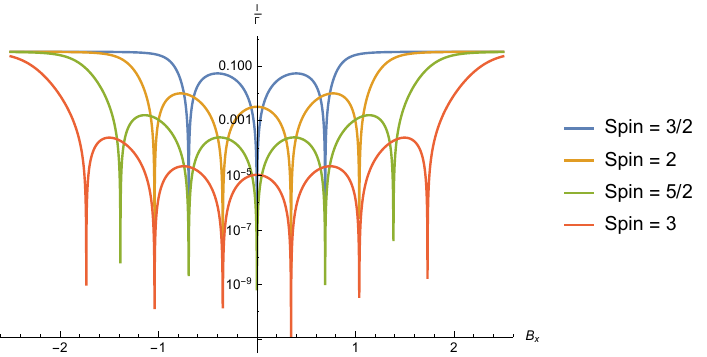}
		\caption{Analytic solution to the Lindblad master equation for SMM's in the Coulomb blocked regime connected to source and drain contacts, under the influence parallel and transverse magnetic field with $s = 3/2$, $s = 2$, $s = 5/2$, $s = 3$, and $K = 0.06$ meV, $K_z = 1$ meV and $\Gamma = 0.38$. }
	    \label{fig:LindbladAnalitica}
\end{figure}

Solving equation (\ref{eq:lindblad_smm}) for the steady state $\left( d\rho / dt = 0 \right)$, we get the following value for the stationary current

\begin{equation}
    \left< I \right>= \Gamma \dfrac{4\Delta E^2}{\Gamma^2 +12\Delta E^2}
    \label{eq:curr1}
\end{equation}
We have to point out that in all our calculations, we used atomic units ($\hbar = e = 1$). From the result given in equation (\ref{eq:curr1}) we see that the stationary current has a Lorentzian form which only depends on the energy splitting, given in equation (\ref{eq09}) and the tunneling transition rate $\Gamma$. In particular, it is possible to see that the current is suppressed when the energy splitting vanishes ($\Delta E = 0$). The vanishing of the energy splitting is the result of the Berry's phase interference, and, as we have previously shown, it depends on the SMM spin(s) directly, i.e. if the spin of the molecule increases the number of dark states increases. In that equation we can also see that if the value of the spin increases then more elements will appear inside the product, and since we are using a perturbative approach with $B_a<<1$,    this implies that $\Delta E$ will get smaller for every new element in the product. This fact is depicted in the intensity of the curves decreasing for bigger spins.  Figure \ref{fig:LindbladAnalitica} shows that the results are consistent with the predictions from our model.

%%%%%%%%%%%%%%%%%%%%%%%%%%%%%%%%%%%%

\section{Numerical Results }

In order to test the accuracy of Eq. \ref{eq:curr1}, we compare the electronic transport predicted by our analytical expression with a numerical solution of the Lindblad equation. Since we have reduced our Hamiltonian to an effective two level Hamiltonian, we can use the QmeQ package \cite{qmeq2017}, which is an open source Python software, to calculate the stationary current driven by differences in chemical potentials between the leads which are tunnel coupled to a quantum dot \cite{qmeq2017}. We chose the QmeQ package since it allows us to introduce contact spin polarization to the leads which is essential to our proposed system in order to detect the Berry phase oscillations in the current. It also introduces parameters such as the Coulomb interaction, the chemical potentials and temperature for the leads among others that are usually neglected in the analytical solutions for this kind of systems.\\

In QmeQ, a system is described by the following general Hamiltonian
\begin{equation}
{\cal H}= {\cal H_{\text{leads}}} + {\cal H_{\text{tunneling}}} + {\cal H_{\text{molecule}}},
\label{eqHqmeq}
\end{equation}
where ${\cal H_{\text{leads}}}$ describes the electrons in the leads, ${\cal H_{\text{molecule}}}$ describes the electrons within the molecule, and ${\cal H_{\text{tunneling}}}$ stands for the tunneling between the leads and the molecule. Although it is possible to use the term ${\cal H_{\text{mol}}}$, to explictly model a magnetic molecule with its associated high spin value, the QmeQ package becomes computationally expensive for these cases. Hence, for our simulations, the electrons within the SMM (${\cal H_{\text{mol}}}$) were described by our effective two level Hamiltonian given in Eq. (\ref{eq082}) plus a term which describes the Coulomb interaction (${\cal H_{\text{Coulomb}}}$).

\begin{equation}
     {\cal H_{\text{molecule}}} = {\cal H}^{q=1}_{\text{single}} + {\cal H_{\text{Coulomb}}}
 \label{e4}
\end{equation}

\begin{equation}
 {\cal H_{\text{Coulomb}}} = \sum_{mnkl}U_{mnkl}d^{\dagger}_n {d}^{\dagger}_m d_k d_l, \; \; \; \;\; m<n.
 \label{e5}
\end{equation}
$U_{mnkl}$ are the Coulomb matrix elements \cite{Goldozian2016}, and $d^\dagger_n$ creates an electron in the state $n$. For the effective two level Hamiltonian we only consider one element that represents the interaction between the two energy levels that were paired up by the presence of the longitudinal magnetic field.
The rest of the terms in the Hamiltonian of Eq. \ref{eqHqmeq}, were obtained by considering two opposite polarized leads which are tunnel coupled to the SMM.

\begin{figure}[htbp]
	    \centering
	    \includegraphics[scale=0.7]{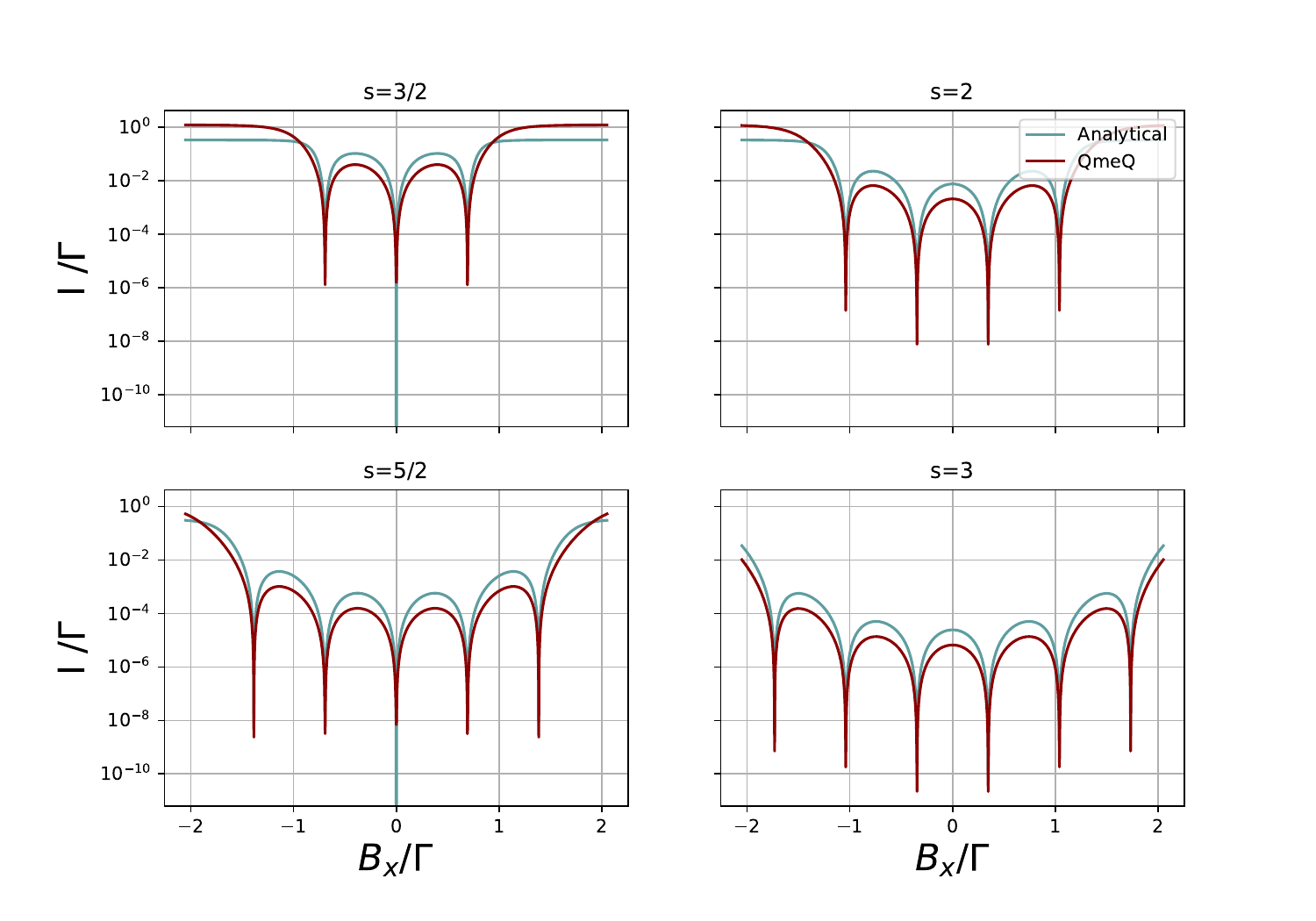}
		\caption{Comparative between the analytical and numerical solutions to the Lindblad master equation for SMMs in the Coulomb blocked regime connected to source and drain contacts, under the influence parallel and transverse magnetic field with $s = 3/2$, $s = 2$, $s = 5/2$, $s = 3$, and $K = 0.06$ meV, $K_z = 1$ meV, $\mu_S = -5$eV, $\mu_D = 5$eV, temperature $t=1\text{x}10^{-5} \text{K}$ and $\Gamma = 0.38$.}.
	    \label{fig:comparativa2}
\end{figure}

Since parameters such as the Coulomb interaction, the chemical potentials for the leads and the temperature of the leads, among others are neglected in the analytical approach (because of its simplicity), in our numerical simulations, we choose a value for the Coulomb interaction according to the chemical potentials from the drain and source leads $\mu_D$ and $\mu_S$, the temperature and bandwidth from the leads and the amplitude $\Gamma$. We found that if $U$ was much bigger than the difference between the chemical potentials in the leads, it was impossible for the current to flow, due to the strong rejection force between the energy levels. On the other hand, the curves tended to turn narrower if the value for $U$ were considerably smaller than the chemical potential gradient. Figure \ref{fig:comparativa2} shows the comparison between the analytical and numerical results, demonstrating the accuracy of the model. The small difference in the curves for the current can be attributed to the numerical parameters that are not involved in the analytical solution, but are specially important in building the system when using QmeQ.

\section{Conclusions}

In summary, we have investigated the electronic transport through SMMs connected to oppositely polarized source and drain contacts using the analytical and numerical solutions of the Lindblad master equation. We introduced an effective Hamiltonian which allows us to have a better understanding of the Berry phase interference as a function of the transverse magnetic field which completely blocks the current flow when we place the single molecule magnet between oppositely polarized leads. Our effective Hamiltonian also captures the number of dark states produced by the Berry's phase interference due to the presence of a transverse magnetic field, which  increases with the SMM's spin, and also shows that the position where they appear depends uniquely on the nature of the molecule. More specifically, on the SMM's anisotropy and unaxial anisotropy (its crystalline nature). We also found that the current's magnitude decreases when the spin increases, due to the product dependence in the splitting of the molecule.\\
Furthermore, by reducing the degrees of freedom we were able to use this effective Hamiltonian in an open source Python software (Qmeq) that allows us to calculate the current through the single molecule magnet with oppositely polarized leads tunnel coupled to the single molecule magnet. The analytical results are well reproduced by our numerical simulations. This opens the door for future research on exploring the quantum transport using the Qmeq software to study the properties of electronic transport through SMMs.

\addcontentsline{toc}{chapter}{BibliografÃ­a}
\bibliographystyle{ieeetr}
\bibliography{bibliografia.bib}

\end{document}